This a pre-print of the accepted manuscript at the International Conference of AI in HCI 2020. Please cite as:
Meske, C. and Bunde, E. (2020). Transparency and Trust in Human-AI-Interaction: The Role of Model-Agnostic Explanations in Computer Vision-Based Decision Support. In: Proceedings of the International Conference of AI in HCI 2020, pp. 54-69.
https://doi.org/10.1007/978-3-030-50334-5_4

# Transparency and Trust in Human-AI-Interaction: The Role of Model-Agnostic Explanations in Computer Vision-Based Decision Support

Christian Meske[0000-0001-5637-9433] and Enrico Bunde[0000-0002-3063-275X]

Freie Universität Berlin, Garystraße 21, 14195 Berlin, Germany
`c.meske@fu-berlin.de`

**Abstract.** Computer Vision, and hence Artificial Intelligence-based extraction of information from images, has increasingly received attention over the last years, for instance in medical diagnostics. While the algorithms' complexity is a reason for their increased performance, it also leads to the 'black box' problem, consequently decreasing trust towards AI. In this regard, "Explainable Artificial Intelligence" (XAI) allows to open that black box and to improve the degree of AI transparency. In this paper, we first discuss the theoretical impact of explainability on trust towards AI, followed by showcasing how the usage of XAI in a health-related setting can look like. More specifically, we show how XAI can be applied to understand *why* Computer Vision, based on deep learning, did or did not detect a disease (malaria) on image data (thin blood smear slide images). Furthermore, we investigate, how XAI can be used to compare the detection strategy of two different deep learning models often used for Computer Vision: Convolutional Neural Network and Multi-Layer Perceptron. Our empirical results show that i) the AI sometimes used questionable or irrelevant data features of an image to detect malaria (even if correctly predicted), and ii) that there may be significant discrepancies in how different deep learning models explain the same prediction. Our theoretical discussion highlights that XAI can support trust in Computer Vision systems, and AI systems in general, especially through an increased understandability and predictability.

## 1 Introduction

The progress in the field of Artificial Intelligence (AI) has led to its wide-spread application in different areas, like the finance or health sector [1-2]. In this context, especially Computer Vision, which refers to machine learning models to extract information from images (e.g., to detect objects), is one of the many research areas in which AI-based systems have achieved high performance or even outperform humans. Already in 2012, a neural network was able to surpass the accuracy of humans when classifying traffic signs [3]. The basis of many breakthroughs in this field was built on the development of deep learning methods. This is a popular branch of machine learning, which simulates structures of the human cerebral cortex and uses large datasets for training and application of multi-layer neural networks [4].

Deep learning is increasingly being examined in the healthcare domain. For example, it can be applied for medical imaging in areas such as radiology (chest radiography), pathology (whole-slide imaging), ophthalmology (diabetic-retinopathy) and dermatology (e.g. skin condition) [5] or parasite detection (malaria) [6-7]. Despite the breakthroughs and progress in this context, one challenge regarding deep learning approaches is its 'black box' characteristic [8]. Due to the high degree of complexity of deep learning-based approaches such as neural networks, there is no inherently comprehensive understanding of the internal processes [9]. AI systems that suffer from this problem are often referred to as opaque [10]. In consequence, there is the trade-off between performance and explainability: while the performance of models increases, the explainability of these approaches decreases [11]. In order to create more transparency, to open the black box and to generate explanations regarding the decisions of AI systems, methods of Explainable Artificial Intelligence (XAI) have been developed. XAI aims to "produce explainable models, while maintaining a high level of learning performance (prediction accuracy); and enable human users to understand, appropriately, trust, and effectively manage the emerging generation of artificially intelligent partners" [12].

In this paper we will focus on XAI and its potential influence on trust. The multi-disciplinary research on trust is conducted, for instance, in philosophy, psychology, sociology, marketing, information systems (IS) or human-computer interaction (HCI) [13-14]. Due to the fact that AI becomes more powerful and is increasingly used in critical situations with potentially severe consequences for humans (e.g., autonomous driving, medical diagnostics), trust towards such systems is an important factor. In the different streams of trust research, there are varying concepts and definitions [13]. We use a concept established by Söllner et al. [14] and thus handle trust as a formative second-order construct.

Our goal is to implement two different neural networks as the basis of a Computer Vision system to detect a disease (malaria) in images (thin blood smear slide images): A Convolutional Neural Network (CNN) and a Multi-Layer Perceptron (MLP). The dataset was obtained from Kaggle and originally stems from the official National Institute of Health. It contains 27,558 images for two classes with 13,779 images for each of the classes 'parasitized' and 'uninfected'. We then aim to generate explanations with the XAI method Local Interpretable Model-Agnostic Explanations (LIME) and use those for the comparison of both neural networks. Overall, we propose the following two research questions. RQ1: How can XAI increase trust in AI-based Computer Vision systems? RQ2: How can XAI methods be used to validate and compare the decision strategy of different AI-based Computer Vision systems?

The paper is structured as follows: First, relevant literature on AI, deep learning and trust is presented. Afterwards, we describe our research design, including the implemented MLP and CNN as well as LIME. This is followed by the results for our implemented neural networks and the generated explanations, and a discussion on the relevance of XAI with respect to trust as well as implications for research and practice. The Paper ends with a conclusion.

## 2 Relevant Literature and Theoretical Background

### 2.1 Artificial Intelligence and Decision Support Systems in Health-Care

AI techniques, especially deep learning models, are increasingly applied in the health sector and fulfil different purposes such as analyzing, interpreting, categorizing, or annotating clinical images [15-16]. Because of the advancements of such AI systems, innovations such as AI-based decision support systems (DSS) for all organizations in general, and especially for health care providers or even as apps for private individuals, are increasingly developed [17-18]. Therefore, it can be stated, that the role of technological decision support in health-care increased [19]. Especially the role of AI is gaining importance, as it is able to integrate various datatypes, which will be used to produce predictive models. Yet, the data collection is a complex process [20-21]. Another reason for the growing interest in AI is based on its performance for different applications. AI was examined in the context of healthcare and DSS with different focuses. For example, machine learning approaches were investigated for predicting the outcome of individual cancer patients, and can help to improve personalized medicine [22]. Another case, where AI has been investigated, is the detection of autism spectrum disorder, which is usually based on behavioral observations, yet there are different approaches to use AI algorithms for detection in data [23]. Moreover, AI-based approaches are investigated for the detection of diabetes and prediction of blood glucose [24]. AI is also being applied for the detection and supervision of illnesses like Parkinson's disease [25] or the diagnosis of asthma [26]. Additionally, such advanced analytics can be implemented to assess whether patients have taken the medications as prescribed or to improve the adherence [27]. Possible benefits from AI for DSS in the healthcare context could include disburden professionals from repetitive tasks, enable timely reaction to critical situations, and to reduce costs, time as well as medical error [27-28]. Decision support systems in general can hence be described as "[…] one of the greatest potential benefits of a digital health care ecosystem." ([21], p. 1).

### 2.2 Computer Vision and Artificial Neural Networks

Computer Vision is a discipline, where deep learning models have helped to significantly increase accuracy [29]. For instance, in the health-care sector, AI-based image interpretation is a well-researched task within medical imaging. There are further areas of application such as image denoising, auto segmentation or image reconstruction [30]. Within the health context there are different image types that are being investigated, whereby diagnostic images are by far the most used health data type [31]. Further concrete application examples of deep learning and computer vision in the health context are the examination of abnormal findings in retinal fundus images [32], recognition of skin conditions such as skin cancer [33] or in the context of neuroscience, the detection of Alzheimer's disease through medical image classification [34]. In our work, we focus on two specific types of neural networks in a Computer Vision system: MLP and CNN. Both neural networks can be categorized as deep learning approaches, whereby deep learning itself is a sub-category of machine learning [34]. Artificial neural networks are inspired by the biological neural network of mammalians. The functional unit of this network is the perceptron, which partitions the input data in separate categories [34-35]. The perceptron is an important element for modern neural networks, which today are composed hierarchically into a network [34].

MLP can also be described as the quintessential example for a deep learning model [36]. Today MLPs are often still applied, e.g., for a comparison between neural networks [37]. CNNs present an approach of state-of-the-art neural networks and are frequently applied for image-level diagnostics, which can be justified with the fact that for many tasks they achieve human-level performance [29]. CNNs are generally composed of different layers, i.e. convolutional, pooling and fully connected layers, whereby the convolutional layer is relevant for the identification of patterns, lines or edges [38]. Pooling layers reduce the number of features, which is done through the aggregation of similar or even redundant features [34]. In general, the CNN gathers different representations across the layers, where they learn individual features of the image [39].

### 2.3 Explainable Artificial Intelligence

The high accuracy of AI has not only been achieved due to an increased performance of hardware but also because of increasingly complex algorithms as used in deep learning approaches. There is hence a trade-off between performance and explainability [11]. Consequently, one of the major issues with AI for DSSs lies in the problem, that they are perceived as black boxes, even by developers. This problematic circumstance hinders the adoption of AI by different stakeholders, for instance due to concerns regarding ethical and responsible clinical implementation of DSSs [21]. For instance, decision trees achieve a rather low performance, yet a high degree of explainability, in contrast to more sophisticated approaches such as neural networks, which can reach a high performance, yet they show a rather low degree of explainability [12]. To solve these problems and to allow for more transparency, methods of "Explainable Artificial Intelligence" (XAI) are developed. The aim of XAI research can be described as to make AI systems more intelligible and human-understandable, which hence become more transparent without decreasing their performance [40-41]. The reasons and motivations for the implementation of XAI methods can be manifold. They can help to increase trust of the user, to better understand and validate the AI systems, to comply with regulations such as the General Data Protection Regulation, and also have an impact on the compliance behavior of employees [42-43]. XAI as a research area has hence a lot of potential to increase trust in AI-based decisions and the underlying algorithms, yet brings new challenges with it, such as what a trustworthy explanation should look like [40]. In literature (e.g., [40]) there are different overarching objectives for XAI: explain to justify (or as we would call it, explain to 'comply'), explain to control, explain to improve and explain to discover (which we would call explain to 'learn' *about* and *from* the system). In addition, so we argue, the goal to comply and to control AI are interconnected, as are the goals to learn and to improve. Eventually, so we argue, the four goals allow individuals and organizations to achieve the overriding objective of *managing* AI. A summary of XAI objectives is depicted in the following Figure 1.

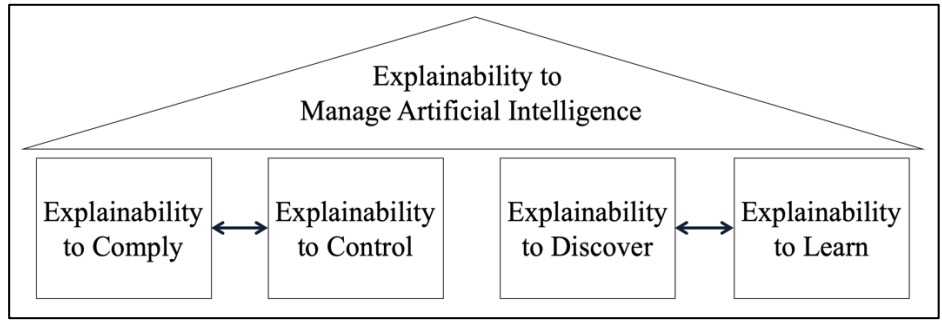

**Fig. 1.** Objectives of Explainable Artificial Intelligence (XAI).

There are numerous overview papers, which establish different categories for the various XAI methods (e.g. [40, 44-45]). For our study, we decided to apply the XAI method Local Interpretable Model-Agnostic Explanations (LIME) as described in more detail in section 4.2 of the research design.

## 3 Theoretical Background: Trust and Human-Computer Interaction

Currently, we can observe a digital transformation of workplaces [46]. In this context, trust is an important component and influences if or how, for instance, AI-based systems will be adopted [47, 44-45]. Especially with regard to critical applications of AI such as for autonomous driving or medical diagnostics, trust plays a major role [48-49]. There are additioan reasons why it is necessary to investigate trust [50]. For example, the risk or the uncertainty associated with a technological interaction can be reduced [14] or the experience with a technology can be created more positive and meaningful [51]. Trust is defined as "[…] the willingness of a party [trustor] to be vulnerable to the actions of another party [trustee] based on the expectation that the other will perform a particular action important to the trustor, irrespective of the ability to monitor or control that other party." ([52], p. 712, cited in [14]). We adapt two possible roles of IT artifacts [14] and apply them to the relationship between a human user and an explanation interface (IT artifact): the explanation interface has the role of the trustee, whereas the human is the trustor. Another role for the explanation interface is the mediator role between human users, who are again the trustors, and the AI system as the trustee (visualized in Figure 2).

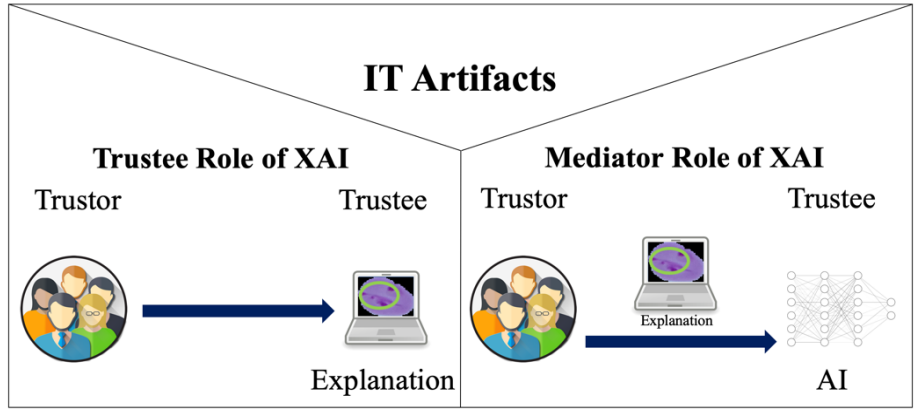

**Fig. 2.** Two Roles of XAI and Explanation Interfaces in Trust Research (modified from [14]).

We are particularly interested how trust towards an explanation or explanation interface can be increased. For the assessment of trust from human users towards an explanation interface, we have adapted the model for trust in IT artifacts, hereinafter referred to as the trust framework [14]. We find this framework suitable for our study, since it is designed for the conceptualization of trust in IT artifacts, which can also represent AI-based Computer Vision systems or explanation interfaces. According to this framework, trust is constituted by the performance, process and purpose of the IT artifact. We are especially interested in the subdimensions of the Process of the IT artifact, on which XAI and explanation interfaces can have an influence: *user authenticity, understandability, predictability, confidentiality, authorized data usage* and *data integrity* (see Figure 3).

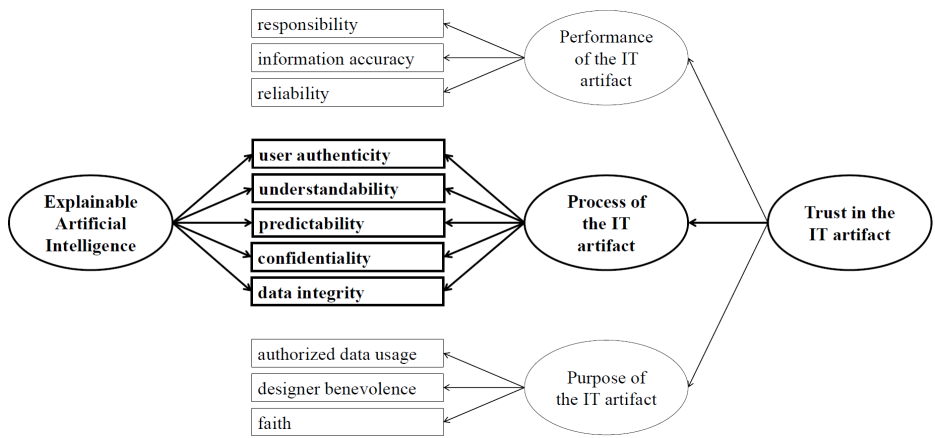

**Fig. 3.** Trust Framework for IT artifacts (modified from [14], p. 7).

We argue, that the explanation interface of an AI system will affect these five formative indicators and hence, influence the trust in the IT artifact. *User authenticity* can be understood as the user's perception that no other user can act unauthorized, in his own name [14]. This is important, for example, when physicians work with an AI-based DSS only themselves or other specific and authorized users should have access to view the prediction or explanation in an interface, access sensible data or even take changes. *Understandability* refers to the fact, that a user understands how the system works, for example, how a (malaria) detection was generated. This point is of high relevance as users want to understand the technology and therefore build more trust [14, 12]. *Predictability* can answer the question how good a user can predict the next actions of the IT artifact [14, 53]. *Confidentiality* refers to the perception of the user that he can control who else is able to access his data, which is related to the indicator understanding [14]. *Data integrity* focuses on the personal data and that they cannot be changed without being noticed, which can be important as users in general want to be in control of their data [14].

## 4 Research Design

### 4.1 Implementing the Multi-Layer Perceptron and Convolutional Neural Networks

Our goal is to train two AI-based Computer Vision models, an MLP and a CNN, to detect malaria in cell images. We then want to use XAI to understand and compare the detection (or 'decision') strategy of each model to increase trust. We have implemented both models with keras and computed the metrics (i.e. accuracy, recall, f1-score) through the scikit-learn classification report. Table 1 provides an overview of the architectures of both deep learning models. As it can be seen, the MLP is a simple multi-layered neural network, while the CNN is inspired by the VGG-16 architecture, whereby we have created a slimmer version here, due to limitations of the computing infrastructure. Furthermore, we have used a batch size of 32, Rectified Linear Unit (ReLu) as activation function, Dropout for regularization, Stochastic gradient descent as optimizer, binary cross entropy as loss function, and a Sigmoid function as last layer activation. The training process would operate for 150 epochs, though we have used early stopping to monitor the validation loss, if it stopped decreasing for 10 epochs, the training was cancelled, and the best weights of the model restored and saved.

**Table 1.** Overview of the Architectures for the MLP and CNN.

| MLP | CNN |
| --- | --- |
| Dense Layer (128, ReLu) | Convolutional Layer (32, 3x3, 1, ReLu) |
| Dense Layer (128, ReLu) | Global Average Pooling (2x2) |
| Dense Layer (128, ReLu) | Convolutional Layer (64, 3x3, 1, ReLu) |
| Dropout (0.5) | Global Average Pooling (2x2) |
| Dense Layer (1, Sigmoid) | Convolutional Layer (128, 3x3, 1, ReLu) |
| | Global Average Pooling (2x2) |

| |
|---|
| Convolutional Layer (256, 3x3, 1, ReLu) |
| Global Average Pooling (2x2) |
| Convolutional Layer (512, 3x3, 1, ReLu) |
| Global Average Pooling (2) |
| Dense Layer (1024, ReLu) |
| Dense Layer (1024, ReLu) |
| Dropout (0.5) |
| Dense Layer (1, Sigmoid) |

### 4.2 Local Interpretable Model-Agnostic Explanations and the Investigated Data Set

The decision to use Local Interpretable Model-Agnostic Explanations (LIME) was made because an XAI method was required, which can be implemented for both models (CNN and MLP). LIME was introduced in 2016 [54] and is also offered as a python library, which simplifies integration into the development environment. In addition, LIME has already been investigated and examined in various tasks such as the classification and explanation of lymph node metastases [55] or recognition of facial expressions [56]. After a few tests, we decided to visualize the two most relevant regions on an explanation for malaria detection. When we had more regions visualized, the problem arose that in part the meaningfulness of the explanation was lost, due to an overload of highlighted regions in the image. Regions that represent the predicted class are highlighted in green (for instance the class: malaria) and regions that stand against the predicted class are highlighted in red (for the class: no malaria).

The dataset was obtained from Kaggle [57] and originally stems from the official National Institute of Health (NIH), which hosts a repository for this dataset [58]. The dataset contains 27,558 images: 13,779 of the class 'parasitized' cell images and 13,779 of the class 'uninfected' cell images. Figure 4 visualizes five randomly selected, exemplary images for both classes. The images of the dataset where of different sizes, so they had to be resized (128x128 pixels). The data was investigated by Rajaraman et al. [6-7] with a focus on the performance of different neural networks. This gives us some comparative metrics, regarding the performance of our own neural networks. Although the focus was not on presenting new benchmarks, it can be argued that performance can also influence the quality of the explanation.

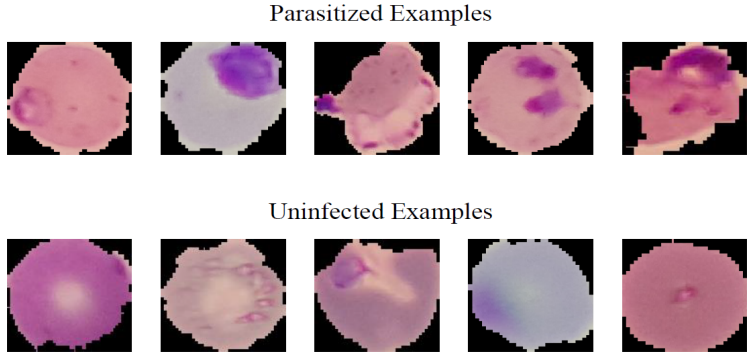

**Fig. 4.** Exemplary images for both classes parasitized (first row) and uninfected (second row).

## 5 Results

### 5.1 Performance of the Computer Vision-Based Malaria-Detection

In the following section we present the performance-related metrics of the artificial neural networks. We will compare the results of the two approaches using the conventional metrics accuracy, recall and f1-score. Rajaraman et al. [6-7] presented benchmark results for different state-of-the-art architectures, such as VGG-16 (accuracy: 95.59%) or VGG-19 (accuracy: 99.09%). Our overall goal was not to exceed these values, yet they can serve as a benchmark. With our own CNN model, we were able to achieve comparable results. Moreover, the CNN has been shown to be a much more powerful and efficient model compared to the MLP. Table 2 gives an overview of the results of the two neural networks, as well as the results achieved for accuracy, recall and the f1-score. Furthermore, the values achieved are shown per class and as a weighted average. The results verify the assumption, that the CNN would outperform the MLP for all metrics.

**Table 2.** Results of the malaria detection based on the CNN and MLP.

| Neural Network | Class | Accuracy | Recall | F1-Score |
|---|---|---|---|---|
| **CNN** | Parasitized | **94.5%** | 97.9% | 96.2% |
|  | Uninfected | **98.1%** | 94.4% | 96.2% |
|  | Weighted Average | **96.3%** | 96.1% | 96.2% |
| **MLP** | Parasitized | 71.0% | 62.0% | 67.0% |
|  | Uninfected | 67.5% | 77.5% | 72.2% |
|  | Weighted Average | 70.2% | 69.8% | 69.6% |

## 5.2 Results of the Application of Explainable Artificial Intelligence

In the exemplary LIMEs, the two most relevant regions are highlighted. If only one region can be seen in an image, it means that the two most relevant regions were next to each other. These can be regions which *support* the decision for its predicted class (green) or which *oppose* the predicted class (red). In Figure 5, four different examples for the parasitized class are depicted. In the first row we see, for example, that the original image contains relevant regions in the lower half of the image. The CNN's explanations are relatively intuitive. For example, (1) a region is highlighted which clearly marks a conspicuous region and a second region, which highlights a mix of conspicuous and inconspicuous areas at the same time. This contrasts with the LIME of the MLP, in which two adjacent regions with two regions lying side by side are marked, which for the most part only include completely irrelevant regions (e.g. (2) and (3))

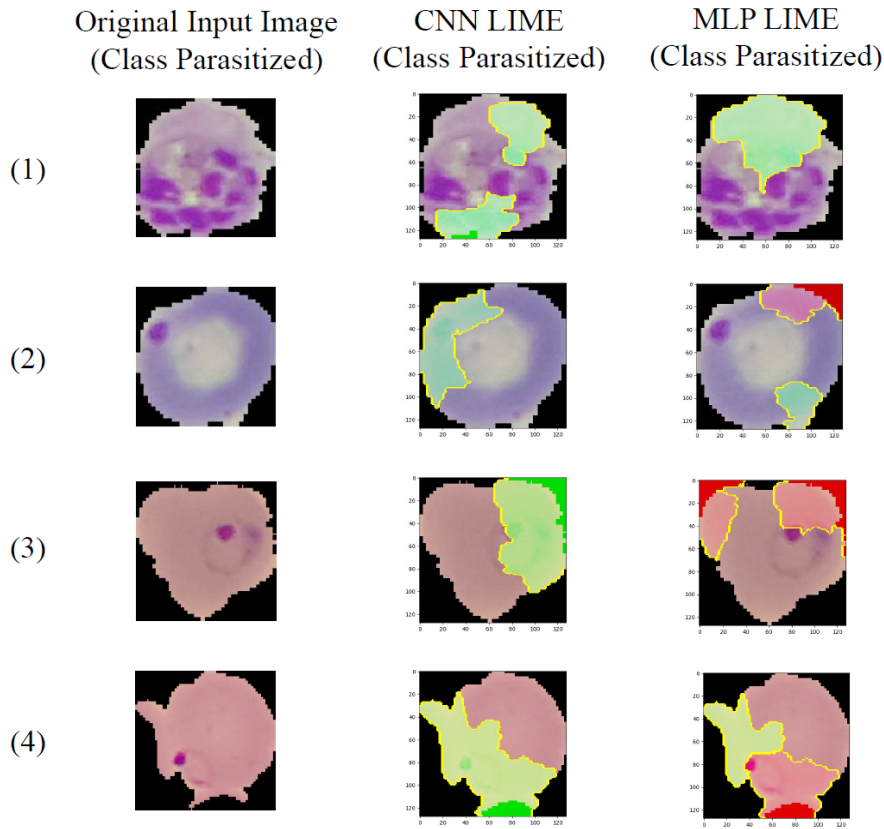

**Fig. 5.** Comparison of LIMEs for the *Correct* AI-Based Classification of *Parasitized* Cells.

Figure 6 shows some LIMEs for the 'uninfected' class. It can be seen again that the CNN correctly highlights regions that stand for the uninfected class, whereby the MLP again highlights regions that may speak for and against the uninfected class. It is interesting that small irregularities in the image are often included in the explanations. For example, this could indicate that the CNN can distinguish the relevant regions from parasitized and uninfected examples, using this ability for classification. Another observation is that in many LIMEs it can be seen that the black borders of the images are often included in the explanation and highlighted as a relevant area, even though this data feature should not play a role for the classification.

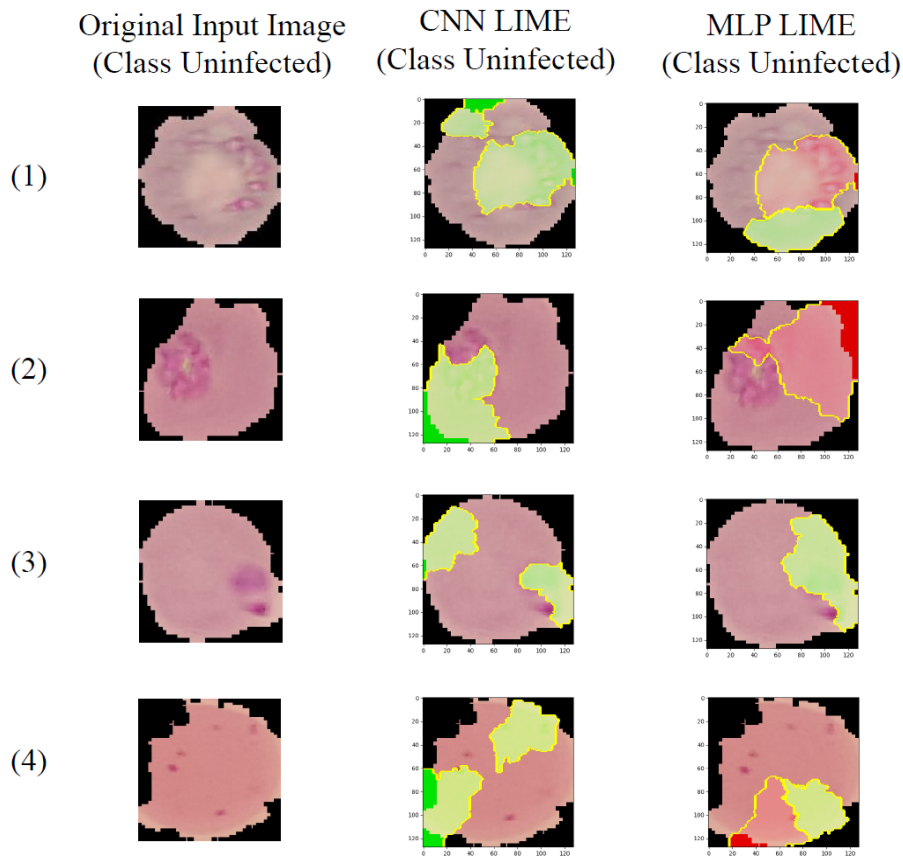

**Fig. 6.** Comparison of LIMEs for the *Correct* AI-based Classification of *Uninfected* Cells.

## 6 Discussion

The evaluation based on the metrics showed that the CNN exceeded the MLP. The CNN was able to achieve more than 96% for all metrics (accuracy, recall, f1-score). These results illustrate how powerful deep learning-based computer vision approaches have become. The results also show that AI-based decision support can be a great support for humans. The better performance of the CNN is also reflected in the LIMEs.

For the most part, the CNN has applied comprehensible decision strategies, detecting relevant features in the cell images, while the MLP often marked irrelevant areas of the cell image, even if correctly classified. An interesting observation was that not all conspicuous regions were highlighted in the LIMEs. In fact, it was more often a mix of relevant and irrelevant regions, which contradicts human expectations and can influence the human-computer trust relationship. Another behavior that can be classified as undesirable behavior is the following. Very often, the black borders of the images were marked as relevant regions in the LIMEs of both models. Yet, they should be unimportant for the classification task.

Based on these findings, in the following we will conceptually discuss and reflect on the adapted trust framework, especially regarding aspects of the *Process of the IT artifact*. To make this discussion more comprehensible, we refer to the following fictitious scenario: a physician implemented a DSS and receives an explanation for a certain prediction, which are presented in an explanation interface.

*User authenticity* plays an important role for the assessment and development of trust. A user (e.g. physician) should be able to be sure that no other user can carry out actions on their behalf, e.g., prescribing medication. This indicator can be transferred to the explanation interface, as it can help to prevent unauthorized persons from accessing it through a personalized login or lock screen. In addition, metadata can be sent for actions that are triggered based on the results in the explanation interface, for example the person who edited data, the time and the device from which an action was initiated, so that user authenticity could be implemented and evaluated in the explanation interface.

*Understandability* is an indicator, which focus directly on the explanation as the goal of XAI: making the results of an AI system more understandable to humans [40]. However, the application scenario, target group and the implemented AI models such as CNN or MLP play a major role here. For complex approaches such as neural networks, there are a variety of XAI methods to open the black box and generate explanations (e.g. LIME) [44-45]. The explanations of certain predictions, also called local explanations in contrast to global explanations regarding the whole AI model, highlight the relevant data features and hence make the decision strategy comprehensible.

*Predictability* is also a relevant indicator, which in our case, is intended to indicate how well a user can use the current explanations to evaluate how the system will handle, for example, new and unknown data. Therefore, the questions 'Why did you do that' or 'Why not something else? should not come up for the user; rather the user should be able to answer these questions himself through the explanation or explanation interface [12].

*Confidentiality* is also linked to the indicator *understandability* [14]: the user wishes to understand how the system works and wants to be in control. In this context, confidentiality refers to questions regarding who else has access to the data or the system. For example, a personalized interface could be created, which is only intended for a specific user and therefore lead to a high degree of confidentiality.

*Data integrity* is similar to the indicator *user authenticity* since this aspect also addresses the explanation interface rather than the sole explanation. It is about the extent to which personal data is processed and that changes to this data should be traceable.

Here, for example, the relevant data could also be displayed in the explanation interface, which was used for the prediction so that the user can see and examine it or even experiment with different data.

## 7  Conclusion

In this study, we investigated how explanations can help to increase trust in AI. Moreover, we were able to demonstrate how to implement XAI to better understand AI in a critical area such as disease detection based on deep learning-based approaches. In doing so, we were able to achieve a certain degree of explainability, which, in addition to the conventional metrics, enabled us to use a further instrument for the comparison of two neural networks. It was also possible for us to increase the explainability without sacrificing performance. We can use the explanations in the form of LIMEs to control the AI's prediction. Based on the visual explanations, we can quickly identify the relevant areas of a predicted class and compare them with our own interpretation of the data and critically reflect on the prediction or decision recommendation. It was also possible to identify a certain level of undesirable behavior, as sometimes areas from the black, irrelevant borders of an image was used to classify malaria. Moreover, a relevant realization was that the mere presentation of an explanation is not be enough for an end-user to evaluate the trustworthiness of an AI. Here, it would be necessary to set up an explanation interface and to augment it with further relevant elements (e.g. the predicted class or confidence).

There are various ways how future research can build on our work. One possibility would be to examine how the quality and performance of the deep learning models can be increased with the help of AI explanations. This could be achieved, for example, by data augmentation (i.e. additional data being generated from the existing data). Moreover, it is still unsolved, how to generate knowledge from AI explanations, or in other words, to learn from what the machine has learned. In addition, it could be examined how the explanations of a CNN differ from those of a Recurrent Neural Network for Computer Vision. Future research should also deal with the evaluation of the adapted trust framework. Another option would be to establish design principles for personalized explanation interface of DSSs, and evaluate those in empirical settings of human-AI interactions.

## References

1. Grace, K., Salvatier, J., Dafoe, A., Zhang, B., Evans, O.: Viewpoint: When Will AI Exceed Human Performance? Evidence from AI Experts. Journal of Artificial Intelligence Research 62, pp. 729-754 (2018).
2. Maedche, A., Legner, C., Benlian, A., Berger, B., Gimpel, H., Hess, T., Hinz, O., Morana, S., Söllner, M.: AI-Based Digital Assistants. Business & Information Systems Engineering 61(4), pp. 535-544 (2019).
3. Ciresan, D., Meier, U., Masci, J., Schmidhuber, J.: Multi-column deep neural network for traffic sign classification. Neural Networks 32, pp. 333-338 (2012).

4. Lu, Y.: Artificial intelligence: a survey on evolution, models, applications and future trends. Journal of Management Analytics 6(1), pp. 1-29 (2019).
5. Kulkarni, S., Seneviratne, N., Baig, M. S., Khan A. H. H.: Artificial Intelligence in Medicine: Where Are We Now? Academic Radiology 27(1), pp. 62-70 (2020).
6. Rajaraman, S., Antani, S. K., Poostchi, M., Silamut, K., Hossai, A., Maude, R. J., Jaeger, S., Thoma, G. R.: Pre-trained convolutional neural networks as feature extractors toward improved malaria parasite detection in thin blood smear images. PeerJ, pp. 1-17 (2018).
7. Rajaraman, S., Jaeger, S., Antani, S. K.: Performance evaluation of deep neural ensembles toward malaria parasite detection in thin-blood smear images. PeerJ, pp. 1-16 (2019).
8. Teso, S., Kersting, K.: Explanatory Interactive Machine Learning. In: Conitzer, V., Hadfield, G., Vallor, S. (eds.) AIES'19: AAAI/ACM Conference on AI, Ethics, and Society, pp. 239-245. Association for Computing Machinery, New York (2019).
9. Schwartz-Ziv, R., Tishby, N.: Opening the blackbox of Deep Neural Networks via Information, (2017), https://arxiv.org/abs/1703.00810, last accessed 2020/01/09.
10. Zednik, C.: Solving the Black Box Problem: A Normative Framework for Explainable Artificial Intelligence, Philosophy & Technology, pp. 1-24 (2019).
11. Gunning, D., Aha, D. W.: DARPA's Explainable Artificial Intelligence (XAI) Program. AI Magazine 40(2), pp. 44-58 (2019).
12. DARPA: Explainable Artificial Intelligence (XAI), DARPA program Update November 2017, pp. 1-36 (2017), https://www.darpa.mil/attachments/XAIProgramUpdate.pdf, last accessed 2020/01/27.
13. Corritore, C. L., Kracher, B., Wiedenbeck, S.: On-line trust: concepts, evolving themes, a model. International Journal of Human-Computer Studies 58(6), pp. 737-758 (2003).
14. Söllner, M., Hoffmann, A., Hoffmann, H., Wacker, A., Leimeister, J. M.: Understanding the Formation of Trust in IT Artifacts. In: George J. F. (eds) In: Proceedings of the 33rd International Conference on Information Systems, ICIS 2012, pp. 1-18 (2012).
15. Jayaraman, P. P., Forkan, A. R. M., Morshed, A., Haghighi, P. D., Kang, Y.-B.: Healthcare 4.0: A review of frontiers in digital health. Wiley Interdisciplinary Reviews-Data Mining and Knowledge Discovery, e1350, pp. 1-23 (2019).
16. Gilbert, F. J., Smye, S. W., Schönlieb, C.-B.: Artificial intelligence in clinical imaging: a health system approach. Clinical Radiology 75(1), pp. 3-6 (2020).
17. Meske, C., Amojo, I.: Social Bots as Initiators for Human Interaction in Enterprise Social Networks. In: Proceedings of the 29th Australasian Conference on Information Systems (ACIS), paper 35, pp. 1-22 (2018).
18. Kemppainen, L., Pikkarainen, M., Hurmelinna-Laukkanen, P., Reponen, J.: Connected Health Innovation: Data Access Challenges in the Interface of AI Companies and Hospitals. Technology Innovation Management Review 9(12), pp. 43-55 (2019).
19. Poncette, A.-S., Meske, C., Mosch, L., Balzer, F.: How to Overcome Barriers for the Implementation of New Information Technologies in Intensive Care Medicine. In: Proceedings of the 21st Human-Computer Interaction International, LNCS vol. 11570, pp. 534-546. Springer, Heidelberg (2019).
20. Stieglitz, S., Meske, C., Roß, B., Mirbabaie, M.: Going Back in Time to Predict the Future – The Complex Role of the Data Collection Period in Social Media Analytics. Information Systems Frontiers (IFD), pp. 1-15 (2018).
21. Walsh, S., de Jong, E. E. C., van Timmeren, J. E., Ibrahim, A., Compter, I., Peerlings, J., Sanduleanu, S., Refaee, T., Keek, S., Larue, R. T. H. M., van Wijk, Y., Even, A. J. G., Jochems, A., Barakat, M. S., Leijenaar, R. T. H., Lambin, P.: Decision Support Systems in Oncology. JCO Clinical Cancer Informatics 3, pp. 1-9 (2019).

22. Ferroni, P., Zanzotto, F. M., Riondino, S., Scarpato, N., Guadagni, F., Roselli, M.: Breast Cancer Prognosis using a Machine Learning Approach. Cancers 11(3), pp. 328:1-9 (2019).
23. Song, D.-Y., Kim, S. Y., Bong, G., Kim, J. M., Yoo, H. J.: The Use of Artificial intelligence in Screening and Diagnosis of Autism Spectrum Disorder: A Literature Review. Journal of the Korean Academy of Child and Adolescent Psychiatry 30(4), pp. 145-152 (2019).
24. Woldaregay, A. Z., Arsand, E., Walderhaug, S., Albers, D., Mamykina, L., Botsis, T., Hartvigsen, G.: Data-driven modeling and prediction of blood glucose dynamics: Machine learning applications in type 1 diabetes. Artificial Intelligence in Medicine, 98, pp. 109-134 (2019).
25. Gil-Martin, M., Montero, J. M., San-Segundo, R.: Parkinson's Disease Detection from Drawing Movements Using Convolutional Neural Networks. Electronics 8(8), pp. 907:1-10 (2019).
26. Spathis, D., Vlamos, P.: Diagnosing asthma and chronic obstructive pulmonary disease with machine learning. Health informatics Journal 25(3), pp. 811-827 (2019).
27. Eggerth, A., Hayn, D., Schreier, G.: Medication management needs information and communications technology-based approaches, including telehealth and artificial intelligence. British journal of Clinical Pharmacology, pp. 1-8 (2019).
28. Khanna, S.: Artificial intelligence: contemporary applications and future compass. International Dental Journal 60(4), pp. 269-272 (2010).
29. Esteva, A., Robicquet, A., Ramsundar, B., Kuleshov, V., DePrisot, M., Chou, K., Cui, C., Corrado, G., Thrun, S., Dean, J.: A guide to deep learning in healthcare. Nature Medicine 25(1), pp. 24-29 (2019).
30. Lewis, S. J., Gandomkar, Z., Brennan, P. C.: Artificial Intelligence in medical imaging practice: looking to the future. Journal of Medical Radiation Sciences 66, pp. 292-295 (2019).
31. Jiang, F., Jiang, Y., Zhi, H., Dong, Y., Li, H., Ma, S. F., Wang, Y., Dong, Q., Shen, H. P., Wang, Y.: Artificial intelligence in healthcare: past, present and future 2(4), pp. 230-243 (2017).
32. Son, J., Shin, J. Y., Kim, H. D., Jung, K.-H., Park, K. H., Park, S. J.: Development and Validation of Deep learning models for Screening Multiple Abnormal Findings in Retinal Fundus Images. Ophthalmology 127(1), pp. 85-94 (2019).
33. Chen, M., Zhou, P., Wu, D., Hu, L. Hassan, M. M. Alamri, A.: AI-Skin: Skin disease recognition based on self-learning and wide data collection through a closed-loop framework. Information Fusion 54, pp. 1-9 (2020).
34. Valliani, A. A., ranti, D., Oermann, E. K.: Deep Learning in Neurology: A Systematic Review. Neurology and Therapy 8(2), pp. 351-365 (2019).
35. Rosenblatt, F.: The Perceptron: A Probabilistic Model For Information Storage and Organization in the Brain. Psychological Review 65(6), pp. 386-408 (1958).
36. Goodfellow, I., Bengio, Y., Courville, A.: Deep Learning, MIT Press (2016).
37. Jang, D.-H., Kim, J., Jo, Y. H., Lee, J. H., Hwang, J. E., Park, S. M., Lee, D. K., Park, I., Kim, D., Chang, H.: Developing neural network models for early detection of cardiac arrest in emergency department. American Journal of Emergency Medicine 38(1), pp. 43-49 (2020).
38. Kim, M., Yun, J., Cho, Y., Shin, K., Jang, R., Bae, H-J., Kim, N.: Deep Learning in Medical Imaging. Neurospine16(4), pp. 657-668 (2019).
39. Saba, L., Biswas, M., Kuppili, V., Godia, E. C. Suri, H. S., Edla, D. R., Omerzu, T., Laird, J. R., Khanna, N. N., Mavrogeni, S., Protogerou, A., Sfikakis, P. P., Viswanathan, V., Kitas, G. D., Nicolaides, A., Gupta, A., Suri, J. S.: The present and future of deep learning in radiology. European journal of Radiology 114, pp. 14-24 (2019).

40. Adadi, A., Berrada, M.: Peeking Inside the Black-Box: A Survey on Explainable Artificial Intelligence (XAI). IEEE Access 6, pp. 52138-52160 (2018).
41. Gunning, D., Stefik, M., Choi, J., Miller, T., Stumpf, S., Yang, G.-Z.: XAI – Explainable artificial intelligence. Robotics 4(37), pp. eaay7120:1-2 (2019).
42. Dosilovic, F. K., Brcic, M., Hlupic, N.: Explainable Artificial Intelligence: A Survey. In: 41st International Convention on Information and Communication Technology, pp. 210-215, Electronics and Microelectronics, Opatija Croatia (2018).
43. Kühl, N., Lobana, J., Meske, C.: Do you comply with AI? Personalized explanations of learning algorithms and their impact on employees' compliance behavior. 40th International Conference on Information Systems (forthcoming), pp. 1-6 (2019).
44. Guidotti, R., Monreale, A., Ruggieri, S., Turini, F., Giannotti, F., Pedreschi, D.: A Survey of Methods for Explaining Black Box Models. ACM Computing Surveys (CSUR) 51(5), pp. 93:1-42 (2018).
45. Ras, G., van Gerven, M., Haselager, P.: Explanation Methods in Deep Learning: Users, values, Concerns and Challenges, 1-15 (2018). ArXiv https://arxiv.org/abs/1803.07517, last accessed 2020/01/27.
46. Meske, C.: Digital Workplace Transformation – On The Role of Self-Determination in the Context of Transforming Work Environments. In: Proceedings of the 27th European Conference on Information Systems (ECIS), pp. 1-18 (2019).
47. Yan, Z., Kantola, R., Zhang, P.: A Research model for Human-Computer Trust Interaction. Proceedings of the 2011 IEEE 10th International Conference on Trust, Security and Privacy in Computing and Communications, pp. 274-281 (2011).
48. Mühl, K., Strauch, C., Grabmaier, C., Reithinger, S., Huckauf, A., Baumann, M.: Get Ready for Being Chauffeured: Passenger's Preferences and Trust While Being Driven by Human Automation. Human Factors, pp. 1-17 (2019).
49. Qasim, A. F., Meziane, F., Aspin, R.: Digital watermarking: Applicability for developing trust in medical imaging workflows state of the art review. Computer Science Review 27, pp. 45-60 (2018).
50. Gulati, S., Sousa, S., Lamas, D.: Design, development and evaluation of a human-computer trust scale. Behaviour & Technology 38(10), pp. 1004-1015 (2019).
51. McKnight, D. H., Carter, M., Thatcher, J. B., Clay, P. F.: Trust in Specific Technology: An Investigation of Its Components and Measures. ACM Transactions on Management Information Systems (TMIS) 2(2), pp. 12-32 (2011).
52. Mayer, R. C., Davis, J. H., Schoorman, F. D.: An Integrative Model of Organizational Trust. Academy of Management Review 20(3), pp. 709-734 (1995).
53. Muir, B. M., Moray, N.: Trust in Automation. Part II. Experimental Studies of Trust and Human Intervention in a Process Control Simulation. Ergonomics 39(3), pp. 429-460 (1996).
54. Ribeiro, M. T., Singh, S., Guestrin, C.: "Why Should I Trust You?" Explaining the Predictions of Any Classifier. Proceedings of the 22nd ACM SIGKDD International Conference on Knowledge Discovery and Data Mining, pp. 1135-1144 (2016).
55. de Sousa, I. P., Vellasco, M. M. B. R., da Silva, E. C.: Local Interpretable Model-Agnostic Explanations for Classification of Lymph Node Metastases. Sensors 19(13), pp. 2969:1-18 (2019).
56. Weitz, K., Hassan, T., Schmid, U., Garbas, J.-U.: Deep-learned faces of pain and emotions: Elucidating the differences of facial expressions with the help of explainable AI methods. TM-Technisches Messen 86(7-8), pp. 404-412 (2019).
57. Kaggle Malaria Cell Images Dataset, https://www.kaggle.com/iarunava/cell-images-for-detecting-malaria, last accessed 2020/27/01.

58. National Library of Medicine – Malaria datasets, https://lhncbc.nlm.nih.gov/publication/pub9932, last accessed 2020/27/01.